# Metric dynamics

S.V.Siparov
State University of Civil Aviation, St. Petersburg, Russian Federation
NRU of Information Technologies, Mechanics and Optics, St. Petersburg, Russian Federation

The suggested approach makes it possible to produce a consistent description of motions of a physical system. It is shown that the concept of force fields defining the systems' dynamics is equivalent to the choice of the corresponding metric of an anisotropic space, which is used for the modeling of physical reality and the processes that take place. The examples from hydrodynamics, electrodynamics, quantum mechanics and theory of gravitation are discussed. This approach makes it possible to get rid of some known paradoxes; it can be also used for the further development of the theory.

**Introduction**

By measuring the motions of a body, the observer classifies them, and may raise the question of their causes. In classical mechanics, the concept of force is introduced as such cause, and the direct proportionality between the resultant force (the forces are considered additive) and the acceleration of the body is postulated (2nd law of dynamics). This suggests the existence of the vector field of forces. Direct measurement of force using a dynamometer is possible only in certain (usually static or stationary) cases. In other cases, the forces and the field strengths, if necessary, are calculated by means of said 2nd law and the measurement of body's acceleration. For this purpose, it is necessary to measure the coordinates of the body in each point of its motion and parameterize the resulting dependency. The natural setting is a real number in the interval of the real axis that is isomorphic to the path between the start and end points of the trajectory corresponding to the body's motion. However, you can also use the time - the number of periods of the periodic steady-state process occurring at the same time, when the body is moving. Mutual influence of the trajectory measurement procedure being performed directly or remotely, and the moving body must be specified. Furthermore, agreement has to be adopted on the rule for calculating the distance with the help of coordinates however they are measured, i.e. on the geometry used in the calculations. After the measurements of coordinates and the corresponding parameter values and after calculating the acceleration, the further agreement on the coupling constant between acceleration and force is needed in order to calculate the force.

In classical mechanics, the following assumptions are used:
- Any environment of the body, or the observer and his instruments for non-contact measurement and observation has no effect on the body - this allows you to take the basic postulate of the existence of inertial reference frames;
- Speed of propagation of information about the coordinates of the body etc. is infinite;
- The geometry of space is Euclidean, and geometric space used in the mathematical modeling of phenomena is a direct sum of the three-dimensional coordinate space and one-dimensional time;
- Coefficient of proportionality between force and acceleration is the inertial mass of the body.

As we know, these assumptions contain arbitraries and contradictions:

- In micro-world, an observer unavoidably affects the object, and in mega-world the object unavoidably affects the observer in a non-deterministic way in both cases; therefore, the postulate of existence of inertial systems in the real physical world (1st law of motion) is not true or requires restrictions;
- Infinite speed of the signal propagation leads to logical paradoxes [1] that make it impossible to have the causality structure in the theory;

- Selection of the modeling space geometry and its dimension is arbitrary and is determined by the observer;
- The nature of the property of inertia, i.e. whether it belongs to the body or to the outside world, can not be regarded as established, and is a philosophical assumption [2].

In modern science these problems are partly overcome, although not always in a completely satisfactory manner. For example, the micro-world is successfully described using quantum mechanics, which, however, contains the paradoxes such as wave-particle duality, the wave function reduction, etc. In mega-world, the successes of general relativity are undeniable, but lately the interpreting of observations confronted with a problem of dark concepts (dark matter and dark energy), the share of which had suddenly become 96% of the material content of the universe, while the carriers of them cannot be found. If we assume that the 1st law of motion dealing with the existence of inertial systems is valid approximately, i.e. starting from some distance between objects, the theory should include a constant with the dimension of distance.

Denying the infinite propagation speed, first, makes it necessary to introduce a universal constant having the dimension of speed in all the theories. And secondly, it pushes to the choice of non-Euclidean geometry of space modeling the physical world in those situations where it is justified. In the simplest case, the SRT uses the 4-dimensional Minkowski geometry, while the GRT uses the 4-dimensional geometry of Riemann. Usage of these geometries allowed to describe the observed effects sparingly and strictly, either no worse than in the corresponding physical theories (compare SRT and Lorentz theory of the electron) or even better (compare GRT and Newton's theory of gravitation). This indicated a new area of application of geometric ideas to the physical world.

As to the problem of inertia, the equivalence principle is used in the theories of macro- and mega-world, so the factor that makes sense of body's mass, sometimes falls out of the equations, which deprives the concept of force of the conceptual significance. In addition, there is a direct micro-world experiment (the so-called Aharonov-Bohm effect, reported in [3]), in which the motion of a particle is determined not by the forces but by potentials.

Thus, the axiomatic basis of classical mechanics (Newtonian dynamics), which allows to compare theoretical results with observations, contains two separate parts, which are not related to each other: the geometry as a way to adequately describe the real space and the processes in it and the force as the cause of the acceleration of bodies.

**1. Geometrization of laws of dynamics**

Both of the above are essentially used in the principle of least action, which is central to physical theory. The equations of Newtonian dynamics that include force fields are identified with the Euler-Lagrange equations that contain the Lagrange function and can be obtained by varying the action functional. An extremal of the action functional coincides [4] with the extremal of the functional of distance between two points (geodesic) in space with the geometry that is chosen to model the real experimental space.

Let us formulate [5] the following propositions:

***Proposition 1***: *The geometry of space, modeling physical reality, is chosen so that the observed free body moves along the geodesic.*

By this, the cause of motion is derived out of consideration, and there is only a description of the observed phenomena using the most appropriate apparatus. At the same time, the source of heuristic ideas that can be used for further development of the theory becomes clearer.

Example 1: In classical mechanics, when there are no forces at a distance (like gravitation), when the information spreads infinitely fast, and when the uniform and rectilinear motion of a free body can be presumably observed, the geodesic is given by

$$\frac{d\vec{r}}{dt} = \vec{v}_0. \tag{1.1}$$

Then, a direct sum of a plane isotropic 3-dimensional space with Euclidean geometry and one-dimensional time can be chosen as the modeling space.

Example 2: If the speed of action propagation is finite, while the free body has an observed acceleration, it is possible to endow the modeling space with due curvature and consider it to be a 4-dimensional space-time with Riemannian geometry, (as is done in GRT). Then the geodesic equation, which coincides with the equation of motion, has the form

$$\frac{dy^i}{ds} + \Gamma^i{}_{kj} y^k y^j = 0, \tag{1.2}$$

and it can be used to determine the metric tensor with the help of the observational data.

Example 3: if the observed accelerated motion of a free body shows an additional dependence on a vector field, then the modeling space should be endowed with anisotropy alongside with curvature. When this occurs, the tangent bundle appears and the geodesic equation becomes more complicated. This situation will be discussed in more detail below.

***Proposition 2***: *The force acting on a body is equal to the product of the matter amount measure (mass) by the acceleration determined by the equation of the geodesic.*

Proposition 2 is formulated so that the force is not the main but the auxiliary concept in contrast to the classical 2nd law of Newtonian dynamics. It is suggested in order to preserve the technical results of the known theory. Relativistic aspects of the concept of (inertial) mass following from the SRT retain their meaning.

***Proposition 3***: *Acceleration of the first body, measured with respect to the second one, is equal in magnitude and opposite in direction to the acceleration of the second body measured with respect to the first one.*

Proposition 3 corresponds to the 3rd Newton's law, which refers to the equality of forces of action and reaction.

The combination of these propositions involves the unification of two parts of the axiomatic basis of theoretical physics, mentioned above, and offers a way to clearly separate the mathematical description of the observed phenomena from their physical essence basing on observations in the real world.

## 2. Geometry of an anisotropic space

Consider a physical situation in which the observed acceleration of a free body depends not only on the coordinates, but also on the velocity vector. If we do not speculate on the causes of the acceleration[1], and try to describe the motion of a body, selecting the geometry of the modeling space, curvature is not enough, and the space must be a tangent bundle.

Let $M = \mathbf{R}^4$ be a differentiable 4-dimensional manifold of class $C^\infty$, $TM$ be its tangent bundle, with coordinates $(x, y) = (x^i, y^i)$; $i = 0 \div 3$. If $c$ is a curve on $M$ defined by $c : [a,b] \to M, t \mapsto (x^i(t))$, then its natural extension to $TM$ is $\tilde{c} : [a,b] \to TM, t \mapsto (x^i(t), y^i(t))$ where $y^i = \frac{\partial x^i}{\partial t}$.

In view of future applications, we note that if we take $x^0 = ct$, where $c = const$ is a constant of the theory, and the parameter $t$ is considered to be time, then $y^0 = c/H = const$, and $H$ is a new constant, arising in connection with the need to measure all the coordinates in the same units, and it has the dimension of frequency. This means that $y^0$ is some fundamental length.

---

[1] Which Newton would not but approve.

The arc length *s*, which is usually taken as a natural parameter on the curve, is $s = \int_0^t \sqrt{g_{ij} y^i y^j} d\tau$, where $g_{ij}$ is metric tensor, and $\sqrt{g_{ij} y^i y^j} \equiv F$. Let the metric tensor depend on *y*, introduced above, i.e. $g_{ij} = g_{ij}(x, y)$. The *x* dependence of the metric tensor will be associated with the curvature of space, and its *y* dependence will mean the anisotropy of space. In other words, in order to calculate the distance between the two points, an additional vector in each of them is needed. In order to ensure constant length of the curve, the metric tensor $g_{ij}$ must be 0-homogeneous with respect to *y*, i.e. $g_{ij}(x, \lambda y) = g_{ij}(x, y); \lambda > 0$, i.e. the metric depends only on the direction of *y*, but not on its magnitude. This is equivalent to $\frac{\partial g_{ij}}{\partial y^k} y^k = 0; i, j, k = 0 \div 3$. (If at the same time $\frac{\partial g_{ij}}{\partial y^k} y^j = 0$, then the metric becomes a Finsler metric [6], but here this case is not considered).

Such metric tensor corresponds to the generalized Lagrange geometry, and it is twice covariant symmetric tensor on *TM* with the following limitations: a) $\det(g_{ij}) \neq 0$ for all (*x*, *y*) on *TM* and b) if the change of coordinates on *TM* is induced by the change of coordinates on *M*, its components transform by the same rule as the components of (0, 2) type tensor on the main manifold *M*. This means that *TM* becomes an eight-dimensional Riemannian manifold, which is an obvious analogy with the six-dimensional phase space known in physics. Four-dimensional base manifold *M* can be viewed as embedded in the eight-dimensional tangent bundle *TM*, and the tangent vectors for all possible curves on *M* form a 4-dimensional vector space, i.e. the tangent space. Tangent bundle can be regarded as isomorphic to the direct sum of the main manifold and the tangent space. The geometry of such eight-dimensional "phase space-time" is quite complicated and requires some specific components (such as non-linear Ehresmann connection).

We restrict ourselves to the simplest case. Let the coordinate transformations $x^i = \Lambda^i{}_j x^j$ be linear, i.e. $\Lambda^i{}_j$ are constants, and the space is considered to be slightly curved and weakly anisotropic with metric $g_{ij} = \eta_{ij} + \varepsilon_{ij}(x, y)$, where $\eta_{ij} = diag\{1, -1, -1, -1\}$ is Minkowski metric on *M*. Moreover, assuming that $\varepsilon_{ij}(x, y) = \sigma \zeta_{ij}(x, y); \sigma << 1$ is small (linearly approximated) anisotropic deformation, the calculations retain only the terms proportional to $\alpha_1 \varepsilon_{ij}, \alpha_2 \frac{\partial \varepsilon_{ij}}{\partial x^k}, \alpha_3 \frac{\partial \varepsilon_{ij}}{\partial y^k}$ and $\alpha_4 \frac{\partial^2 \varepsilon_{ij}}{\partial x^l \partial y^k}$ where all the coefficients $\alpha_k = O(1), \sigma \alpha_k << 1, k = 0 \div 3$. Then the required geometry substantially simplifies. The definition of $y^i$ allows to raise and lower indices corresponding both to "horizontal" (*x*-), and "vertical" (*y*-) values on the tangent bundle, using the same metric tensor $g_{ij}(x, y)$, i.e., use the simplest case – the so-called Sasaki lift [7].

### 3. The equation of motion

In accordance with Proposition 1, the equation of motion is a geodesic of the space chosen for the modeling of physical reality. We define the geodesics as extreme curves for the arc length $s = \int_0^t F d\tau$, where $F = \sqrt{(\eta_{ij} + \varepsilon_{ij}(x, y)) y^i y^j}$. Variation procedure for this expression, which gives the Euler-Lagrange equation in the form $\frac{\partial F}{\partial x^k} - \frac{d}{ds}(\frac{\partial F}{\partial y^k}) = 0$, is given in more detail in [8,9].

With the adopted linear approximation, the equation of geodesic in this anisotropic space has the form

$$\frac{dy^i}{ds} + (\Gamma^i_{lk} + \frac{1}{2}\eta^{it}\frac{\partial^2 \varepsilon_{kl}}{\partial x^j \partial y^t} y^j) y^k y^l = 0, \qquad (3.1)$$

where $\Gamma^i_{jk} = \frac{1}{2}\eta^{ih}(\frac{\partial \varepsilon_{hj}}{\partial x^k} + \frac{\partial \varepsilon_{hk}}{\partial x^j} - \frac{\partial \varepsilon_{jk}}{\partial x^h})$ are the connection coefficients. In this case they are conventional Christoffel symbols, that depend also on *y*. Unlike geodesic equation (1.2) in the Riemann space, the resulting equation (3.1) shows that in anisotropic space the use of coordinate transformation such that all $\Gamma^i_{lk}$ become equal to zero, does not make the system locally inertial. This eliminates the paradox due to the suggestion of the inertial reference frames existence.

Let us list the assumptions that will be used:

1) Under the previous section, only linear terms proportional to $\alpha_1 \varepsilon_{ij}, \alpha_2 \frac{\partial \varepsilon_{ij}}{\partial x^k}, \alpha_3 \frac{\partial \varepsilon_{ij}}{\partial y^k}$ and $\alpha_4 \frac{\partial^2 \varepsilon_{ij}}{\partial x^l \partial y^k}$ wherein $\alpha_n = O(1); i, j, l, k = 0 \div 3; n = 1 \div 4$ are retained in the metric;

2) Components $y^1$, $y^2$, $y^3$ can be neglected in comparison with $y^0$;
3) Time derivative in the equation of a geodesic can be neglected in comparison with the derivatives with respect to coordinates;
4) In the *y*-subspace $y^0$ derivative can be neglected in comparison with the derivatives with respect to $y^1$, $y^2$ and $y^3$.

As a result, equation (3.1) takes the form [9,10]

$$\frac{dy^i}{ds} + \Gamma^i_{00} + \frac{1}{2}\eta^{ik}\frac{\partial^2 \varepsilon_{00}}{\partial x^j \partial y^k} y^j = 0. \qquad (3.2)$$

Thus, it turns out that the only component of the metric tensor, which remains in the equations for the case of small curvature and weak anisotropy, is $\varepsilon_{00}$. Spatial 3D-sectional view of the equation (3.2) has the form

$$\frac{d\vec{v}}{dt} = \frac{c^2}{2}\left\{-\nabla \varepsilon_{00} + \nabla(\vec{v}, \frac{\partial \varepsilon_{00}}{\partial \vec{v}}) + [\vec{v}, rot \frac{\partial \varepsilon_{00}}{\partial \vec{v}}]\right\}, \qquad (3.3)$$

It gives the equation of motion in terms of the metric tensor of the anisotropic space with regard to the velocity of the probe body.

If $\varepsilon_{00} \neq \varepsilon_{00}(y)$, then $\frac{d\vec{v}}{dt} = -\frac{c^2}{2}\nabla \varepsilon_{00}$, and this expression leads to the known geometric formalism of GRT for gravitation, as well as to the formalism of classical mechanics as a whole. If $\varepsilon_{00} \neq \varepsilon_{00}(x)$, and $\frac{c^2}{2} rot \frac{\partial \varepsilon_{00}}{\partial \vec{v}} \equiv \vec{\Omega}$, then the trajectory of a free particle motion is a kind of helix whose axis is directed along vector $\vec{\Omega}$. If at the same time, $\nabla(\vec{v}, \frac{\partial \varepsilon_{00}}{\partial \vec{v}}) = 0$, then the radius

$R$ and the pitch $b$ of such a helix are $R = \dfrac{v \sin(\vec{v},\vec{\Omega})}{|\vec{\Omega}|}$ and $b = 2\pi \dfrac{v \cos(\vec{v},\vec{\Omega})}{|\vec{\Omega}|}$, and the period $\tau$ is equal to $\tau = 2\pi |\vec{\Omega}|^{-1}$. Here, if $\vec{v}$ initially lies in a plane perpendicular to $\vec{\Omega}$, then $b = 0$, the free motion is finite, and the particle stays in a plane and moves in a circle of radius $R$. If vector $\vec{v}$ is initially collinear to $\vec{\Omega}$, then $\vec{v}$ remains constant. Helix is a basic form of the trajectory of the particle's free motion in this anisotropic space.

If $\varepsilon_{00}$ component of the metric tensor become more complex, the helix axis ceases to be a straight line. If the metric is such that this axis presents a closed circle of radius $R_M$, which is greater than or equal to twice the radius $R$ of the helix, then the free motion takes place on the surface of a torus. If

$$n_1 R_M = n_2 R, \qquad (3.4)$$

wherein $n_1$ and $n_2$ are integers, the trajectory on the torus surface becomes closed, and since in this case $n_1 2\pi R_M = n_2 b$, $\tan(\vec{v},\vec{\Omega})$ is a rational number. The typical linear dimension of the body, whose surface is a location of a particle free path, is equal or larger than $2R$.

### 4. The field equations

In classical physics, they talk about the material carriers of the cause acting on a moving body and associate them with the appropriate (force, physical) fields that can be measured. However, the direct measurement of the forces can only be performed in certain (static or stationary) cases, and usually the forces are calculated using the law of classical dynamics. In geometrodynamics (i.e., in GRT) the role of this "cause" is given to the geometrical properties of the modeling space used to describe the body's motion. At the same time, the derivation of the field equations by the variation principle is based on the notion of the Lagrangian, the choice of which is a kind of art and is implicitly associated with the classical dynamics of forces and relevant concepts. In frames of the geometrization of dynamics, which is discussed in this paper and is based on the observed motions, it is natural to preserve the geometric approach, but the pass to the physical interpretation of the obtained relations should be performed only at the last stage, depending on the class of problems under consideration.

In the anisotropic space, every point is equipped with a vector. Let us note some mathematical circumstances related to the presence of an arbitrary vector field. Consider an arbitrary covariant vector with components $B_k = (B_0, \vec{B}), \vec{B} = (B_1, B_2, B_3)$ and construct an antisymmetric covariant tensor, $F_{ik} = B_{k,i} - B_{i,k}$, where $B_{k,i} = \dfrac{\partial B_k}{\partial x^i}$. Then the expression

$$\frac{\partial F_{ij}}{\partial x^k} + \frac{\partial F_{jk}}{\partial x^i} + \frac{\partial F_{ki}}{\partial x^j} = 0 \qquad (4.1)$$

is a geometric identity, sometimes called the Maxwell identity, valid for any geometry.

Writing this identity in the components [9,11], one can introduce a formal notation for a pair of new vectors constructed out of the components of tensor $F_{ik}$

$$\vec{F}^{(*)} = (F_{12}, F_{31}, F_{10}); \vec{F}^{(**)} = (-F_{30}, F_{20}, -F_{23}) \qquad (4.2)$$

Then we get a couple of homogeneous equations

$$\frac{\partial \vec{F}^{(**)}}{\partial t} + rot\vec{F}^{(*)} = 0,$$

$$div\vec{F}^{(**)} = 0$$

(4.3)

where $t \equiv x^0$. The type of geometry of the space has no effect on these equations.

Let us arbitrarily choose and fix the geometry of space: let it correspond to the metric tensor $g^{ik}$, which makes it possible to switch from the covariant components of the tensor $F_{ik}$ to its contravariant components $F^{ik}$ by the formula, $F^{ij} = g^{ik} g^{jm} F_{mk}$. We introduce new additional symbols, namely, a contravariant 4-vector, $I^i = (I^0, \underset{\rightarrow}{j}), \underset{\rightarrow}{j} = (I^1, I^2, I^3)$, such that $I^i = \frac{\partial F^{ij}}{\partial x^j}$. Using the notation

$$I^0 \equiv \rho; I^1 \equiv j_x; I^2 \equiv j_y; I^3 \equiv j_z,$$

(4.4)

get the second pair of – now inhomogeneous – equations

$$rot\vec{F}^{(**)} - \frac{\partial \vec{F}^{(*)}}{\partial t} = \underset{\rightarrow}{j}.$$

$$div\vec{F}^{(*)} = \rho$$

(4.5)

Never going beyond the mathematical formalism, we call $\rho$ the density of sources distribution, and call $\underset{\rightarrow}{j}$ the current density. In order to describe these quantities present in equations (4.3, 4.5), we can formally apply the appropriate integral theorems. This will lead to the expression known as the continuity equation

$$div\underset{\rightarrow}{j} + \frac{\partial \rho}{\partial t} = 0.$$

(4.6)

The relations (4.3, 4.5, 4.6) are always true, because they are the consequences of the mathematical identity (4.1).

If vectors $\vec{F}^{(*)}$, and $\vec{F}^{(**)}$, composed out of the components of tensor $F_{ik} = B_{k,i} - B_{i,k}$, do not depend on $t$, then it follows from the equations (4.3, 4.5) that $\vec{F}^{(*)} = -\nabla \varphi$, where $\varphi$ is a certain scalar function (*scalar potential*) such that it satisfies the Poisson equation

$$\Delta \varphi = -\rho,$$

(4.7)

and, in the absence of sources ($\rho = 0$), it satisfies the Laplace equation

$$\Delta \varphi = 0.$$

(4.8)

For a point source, $\rho = q\delta(r)$, the introduced scalar potential satisfies the expression

$$\varphi \sim \frac{q}{r}.$$

(4.9)

Equation (4.7) is linear, so, for the distributed sources, the superposition principle yields

$$\vec{F}^{(*)} = -\nabla(\int \frac{\rho(r)}{|\vec{r} - \vec{r}_0|} dV) \tag{4.10}$$

If the sources distribution is finite, homogeneous and has a spherical symmetry, then for the exterior problem we come to the formula (4.9), where $q = \int_V \rho dV$, and for the interior problem the expression (4.10) leads to

$$\varphi \sim q_1 r^2 \tag{4.11}$$

where $q_1 = \int_{V_1} \rho dV$, and $V_1$ is the volume of a sphere of radius $r$.

It turns out that vector $\vec{F}^{(**)}$ can be represented as $\vec{F}^{(**)} = rot\vec{D}$, where $\vec{D}$ is another new vector – the so-called (3-dimensional) *vector potential* which satisfies the equation

$$\Delta \vec{D} = -\vec{j}, \tag{4.12}$$

and, correspondingly,

$$\vec{F}^{(**)} = rot(\int \frac{\vec{j}(r)}{|\vec{r} - \vec{r}_0|} dV). \tag{4.13}$$

The 4-vector $(\varphi, \vec{D}) \equiv U_k$ characterizes the geometric properties of the anisotropic space, each point of which is equipped with a given vector, $B_k = g_{ik} B^i$. If vector $B^i$ is the 4-velocity of the probe body, $y^i = \frac{\partial x^i}{\partial t}$, all the above arguments remain valid.

If there are no sources and currents, i.e. $\rho = 0; \vec{j} = 0$, then the system of equations (4.5) still has nonzero solutions that satisfy the equation

$$g^{ki} \frac{\partial^2 U^i}{\partial x^k \partial x^l} = 0. \tag{4.14}$$

For the Minkowski metric, $g_{ik} \equiv \eta_{ik}$, equation (4.14) represents the *wave equation* written in the four-dimensional form. For other geometries, the resulting expressions may be also of interest. In particular, for the case considered in this paper, $g_{ik} = \eta_{ik} + \varepsilon_{ik}(x, y)$, and the wave equation becomes inhomogeneous. Since the correction is linear in $\varepsilon$, it must be preserved in calculations. However, if $U^i$ is also proportional to $\varepsilon$, then the correction can be neglected, and the equation (4.14) will be the usual wave equation.

Pithiness and ease of the obtained formal expressions makes them attractive for use in the interpretations of the observed physical phenomena. In particular, the wave motion is well known in nature and, thus, it can be adequately described in terms of the mentioned potentials. Besides, the measurements of some observed interactions for the certain scales and systems of sources suggest the dependences of the corresponding physical fields proportional to ~ *1/r²* and to ~ *r*. These physical fields can be also described with the help of imposed potentials using conventional operations and corresponding designations.

Thus, the considerations associated with the geometric identity (4.1) lead to the following conclusions:

1. No matter what is the nature of an arbitrary vector field which is used in the modeling of physical phenomena in real space (be it a physical field or a purely mathematical object), it is equivalent to the existence of "Coulomb potential" and to wave processes that determine the motion of a probe body;
2. The geometry chosen to describe the situation does not affect the scalar potential and always gives a "Coulomb's law", but it reveals itself in the description of wave processes [9];

This also means that during the experimental measurements, certain attention should be paid to the establishment of a geometric nature (tensor rank) of the objects involved into the mathematical description. That is, if we measure a "physical field" and consider it a vector, if we observe a source and consider it singular, it will inevitably lead to the inverse square law by virtue of eq. (4.1). And if the sources are continuously distributed in a finite domain with sufficiently smooth boundary, then, far outside the domain, the field will remain the same (in general relativity this is called an "isle model"), and inside the domain the field will be linearly dependent on the distance from a fixed point, the field in which is supposed to be given (see [9] for details).

Besides:

3. The vector field corresponding to the velocity of a probe body can be also regarded as the vector field in question;
4. If the vector field includes the metric tensor, then the corresponding part of the metric has wave character.

Notice, that the known *gravitational* field equations, obtained by Hilbert and Einstein with the help of variation principle, contain the energy-momentum *tensor*. In their derivation, a different geometric identity was used, namely, the integral divergence theorem. This fact serves as a link between the basic axioms mentioned in the Introduction. However, it is clear that within the above three Propositions, the second of the axioms (the reason of acceleration or rather, the way to describe it) is not independent of the first, so the object of the variation procedure should be changed.

**5. Several notes on the canonical formalism**

In the study of the causes of motion of physical systems in frames of metric dynamics, the formulation of the least action principle should be changed. Here it will be used in the following form: for any physical system, there exists the geometry of the modeling space such that the integral, called action, reaches minimum for the actual motion of the system. This suggests that in the geometric approach, the meaning of a number of concepts of the canonical formalism and their properties become different from the classical ones. Nevertheless, writing them down is still meaningful.

Since the equations of motion are already given, we have to go the opposite way and find the structural elements characteristic of the traditional variation approach. The functional of action is the integral of the Lagrangian *L*, which in classical mechanics is the difference between the kinetic and potential energies *T* and *U*. Let $[\vec{v}, rot \frac{\partial \varepsilon_{00}}{\partial \vec{v}}] \equiv \nabla \varphi_{(a)}$ and call anisotropic potential $\Phi_{(a)}$ a value determined by the formula

$$\Phi_{(a)} = \frac{c^2}{2}\left\{\varepsilon_{00} - (\vec{v}, \frac{\partial \varepsilon_{00}}{\partial \vec{v}}) - \varphi_{(a)}\right\}. \tag{5.1}$$

Then the motion of a free body in a slightly curved and weakly anisotropic space takes place in accordance with the equation

$$\frac{d\vec{v}}{dt} = -\nabla \Phi_{(a)} \tag{5.2}$$

Since the geodesic equation (3.1), (3.2) or (3.3) is now the equation of motion, it can be considered the Euler-Lagrange equation, $\frac{d}{dt}\frac{\partial L}{\partial \vec{v}} = \frac{\partial L}{\partial \vec{r}}$. We call the kinetic and potential energies $T$ and $U$ the expressions that have the forms

$$T = \frac{mv^2}{2}\left\{\sin^2(\vec{v},\vec{\Omega}) + \cos^2(\vec{v},\vec{\Omega})\right\}, \tag{5.3}$$

$$U = \frac{mc^2}{2}\left\{\varepsilon_{00} - (\vec{v}, \frac{\partial \varepsilon_{00}}{\partial \vec{v}}) - \varphi_{(a)}\right\}, \tag{5.4}$$

where $\frac{c^2}{2} rot \frac{\partial \varepsilon_{00}}{\partial \vec{v}} \equiv \vec{\Omega}$. If $\varepsilon \neq \varepsilon(y)$, the expression (5.3) takes the usual form of kinetic energy, and the expression (5.4) can be related to the potential energy in general relativity. In this case, according to Proposition 2, a formal expression for the force $\vec{F}_{(a)}$ obtained by multiplying both sides of the equation (3.3) by the mass of the probe body, $m$ will have a familiar look, $\vec{F}_{(a)} = -m\nabla \Phi_{(a)}$.

In metric dynamics, we will call closed such a system, for which the change in the number of its parts does not require changes in the geometry used to describe its motion. Lagrange function $L = T - U$ of such a system is additive, i.e. it is the sum of the Lagrangians of the system parts. The interaction of these parts (in particular, belonging to the outside world), considered in classical mechanics, in metric dynamics is presented by the variations of geometry and is accompanied by changes both in expression (5.3) and (5.4). The law of conservation of the total energy of a closed system is preserved, since a parameter on the segment isomorphic to the body's trajectory is distributed uniformly. The total momentum of a closed system consisting of $N$ bodies and given by

$$\vec{P} = \sum_n \frac{\partial L}{\partial \vec{v}_n}; n = 1,2...N \tag{5.5}$$

acquires an additional geometric meaning and is also preserved, although it is determined by a more complicated expression than usual, for example, despite zero velocity value, the momentum can be still non-zero, and in this case it can be called rest momentum.

Note that the fundamental mechanical motion in an anisotropic space has a helical form; this is why there is no need to talk about the *analogy* between mechanics and wave (geometric) optics. This means that Fermat's principle of minimal phase change between the start and end points of wave propagation is now directly applicable to the motion of the probe body. The first term in eq. (5.3) is related to the translational motion of a particle along the axis $\vec{\Omega}$, and the second term is related to the rotational motion around this axis. Therefore, if $\vec{\Omega} \neq 0$ the kinetic energy for $b = 0$ does not vanish. The energy in this case can be interpreted as rest energy[2]. In this case, the rest momentum would be equal to the product, $\vec{p} = mR\vec{\Omega}$. When a particle moves along the helix, it can be regarded as "suffering the action of a quasi-Lorentz force" which performs no *work*. For this reason, a new canonical concept associated with the period of the helix and the oriented volume can be introduced. In classical theory, it is relevant to the action-angle variables and to the adiabatic invariant [12].

Equations (4.3) and (4.5) obtained in the previous section with the help of geometrical identity (4.1) can be also considered as the equations of motion, but for the other generalized coordinates. They can be obtained by varying a certain Lagrangian, but by the other variables. In particular, if we label

$$T_f = \int \left( (\vec{F}^{(*)})^2 - (\vec{F}^{(**)})^2 \right) dV , \tag{5.6}$$

and call the *energy* already this (geometric) expression, then after its variation, the equations (4.3) and (4.5) will be obtained. Thus, the use of anisotropic space to model the real physical space coupled with the onset phenomena, upholds the formal mathematical procedures related to the canonical equations, but gives the used concepts a somewhat different sense (see also [13]).

## 6. Hydrodynamics and Electrodynamics

Let us use the notation common in hydrodynamics. We introduce $u_i \equiv \dfrac{c^2}{2} \dfrac{\partial \varepsilon_{00}}{\partial y^i}$, where $c$ is a fundamental parameter with the dimension of speed, and take $\Omega_i \equiv \dfrac{1}{2} rot u_i$ and $\varphi = \dfrac{c^2}{2} \varepsilon_{00} - (\vec{v}, \vec{u})$. The relationship of these notations with symbols $\vec{B}, \vec{F}^{(*)}, \vec{F}^{(**)}$ used above is obvious. Let us regroup equation (3.3), and multiply and divide its right hand side by a constant value, $\rho$. Then it takes the form

$$\frac{\partial \vec{v}}{\partial t} + (\vec{v}, \nabla)\vec{u} + 2[\vec{\Omega}, \vec{v}] = -\frac{1}{\rho} \nabla \left\{ \frac{\rho c^2}{2} \varepsilon_{00} - (\vec{v}, \vec{u})\rho \right\} \tag{6.1}$$

We assume vector $\vec{u}$ characterizing the anisotropy of the modeling space, be the flow rate of an incompressible fluid. We assume the value of $\rho$ be a constant density of the fluid in the neighborhood of the probe body (which is "liquid particle"), $\rho = const$. Then, assuming that the gradient of the two terms on the right side is the effective pressure, we obtain the Euler equation for a (rather large) region of the liquid rotating as a whole with angular velocity, $\vec{\Omega}$. The meaning of equation (4.6) is also evident.

If necessary, repeat the calculations for the components of $F_{ik}$ and get for hydrodynamics the expressions, similar to eqs. (4.3, 4.5), for the case when $\vec{u}$ is considered the flow rate of an incompressible fluid. Accordingly, equations (6.1) have both potential and wave solutions. Indeed, considering the speed of the liquid particle equal to the speed of its environment, $\vec{v} \cong \vec{u}$, we seek a solution of eq. (6.1) in the form

$$\vec{v} = \vec{V} e^{i((\vec{k}, \vec{r}) - \omega t)} . \tag{6.2}$$

---

[2] In molecular-kinetic theory it will be related to the absolute zero of temperature.

The amplitude $\vec{V}$ is assumed to be small enough to neglect the term, $(\vec{v},\nabla)\vec{u} \to 0$. Then

$$\frac{\partial \vec{v}}{\partial t} + 2[\vec{\Omega},\vec{v}] = -\nabla \varphi. \tag{6.3}$$

Select a local $Oz$ axis parallel to vector, $\vec{\Omega}$, and apply the *rot* operator to both sides of eq. (6.3), given that $rot[\vec{\Omega},\vec{v}] = \vec{\Omega} div\vec{v} - (\vec{\Omega},\nabla)\vec{v} = -(\vec{\Omega},\nabla)\vec{v}$. Then, we obtain [14]:

$$\frac{\partial}{\partial t} rot\vec{v} = 2\Omega \frac{\partial \vec{v}}{\partial z} \tag{6.4}$$

Dispersion equation has the form

$$\omega = 2\Omega \frac{k_z}{k} = 2\Omega \cos\theta; \theta = (\vec{k} \wedge \vec{\Omega}), \tag{6.5}$$

and the velocity of the wave is given by

$$\vec{U} = \frac{\partial \omega}{\partial \vec{k}} = \frac{2\Omega}{k}\{\vec{v} - \vec{n}(\vec{n},\vec{v})\}; \vec{v} = \frac{\vec{\Omega}}{\Omega}; \vec{n} = \frac{\vec{k}}{k}$$

$$U = \frac{2\Omega}{k}\sin\theta \tag{6.6}$$

In such a wave, the velocity vector, $\vec{v}$ of the fluid particle retains its value and varies only in direction. In classical hydrodynamics, such waves are generated by "Coriolis force" and are called inertial waves; they have no dependence on such force characteristics as pressure. However, in the geometric theory, where both the original geodesic equation and Maxwell's identity are present, there is no Coriolis force, but there is only the metric tensor describing the anisotropic space, in which the motion of bodies takes place. In this space, the zero component of metric is the wave in accordance with the condition $\vec{v} \cong \vec{u}$, i.e. $\varepsilon_{00} = \frac{2V^2}{c^2}e^{2i((\vec{k},\vec{r})-\omega t)}$.

Let us now designate, $\vec{F}^{(*)} \equiv \vec{E}$, $\vec{F}^{(**)} \equiv \vec{H}$ and interpret them as the stresses of electric and magnetic fields. Then, the equations (4.3, 4.5) are usual Maxwell equations and the corresponding stresses are determined by formulas (4.10, 4.13). Using the notation

$$\frac{c^2}{2}\varepsilon_{00} \equiv \varphi \;, \quad \frac{c^2}{2}\frac{\partial \varepsilon_{00}}{\partial \vec{v}} \equiv \frac{1}{c}\vec{A}, \tag{6.7}$$

the equation of motion of a probe body takes the familiar form

$$\frac{d}{dt}\vec{v} = \left\{-\nabla \varphi + \frac{1}{c}[\vec{v}, rot\vec{A}] + \frac{1}{c}(\vec{v}\cdot\nabla)\vec{A}\right\} \tag{6.8}$$

Here, the values of $\varphi$ and $\vec{A}$ are so-called scalar and vector potentials of the *electromagnetic field*. The meaning of the continuity equation (4.6) is obvious. As is known from electrodynamics (and mentioned above), in the particular case of $\varphi \neq \varphi(r); rot\vec{A} = const$ a charged particle moves along a helix, whose axis is directed along vector, $rot\vec{A}$. And electromagnetic waves are classic result of theoretical physics, predicted by Maxwell on the base of the equations (4.3) and (4.5) and on the corresponding interpretation. In particular, $\vec{A} = \mathrm{Re}\{\vec{A}_0 e^{i(\vec{k}\vec{r} - \omega t)}\}$, where $\vec{k} = \dfrac{\omega}{c}\vec{n}$ is the wave vector, $\vec{n}$ is a unit vector in the direction of wave propagation, $\vec{E} = ik\vec{A}$ is the stress of "electric field", $\vec{H} = i[\vec{k}, \vec{A}]$ and is the stress of "magnetic field". Note that the electric field vector rotates in the plane perpendicular to the direction of wave propagation. Charge motion in the wave with circular polarization takes place in the same circumferential plane.

Thus, in the case of hydrodynamics and electrodynamics, the approach under consideration is simply another language for modeling the known observations, leading to almost the same results as the previous one. We can say that instead of the concept of a physical field now the concept of metric field is used, and the physical meaning can be assigned to it not as initially "inherent" but in view of possible interpretation.

### 7. Quantum Mechanics

Characteristic object of a micro-world, whose properties can be studied in quite a variety of ways, is the atom. As follows from experiments, it is a compound dynamic system (planetary model), a direct measurement of whose parameters is hardly possible, and hence, the atom can be also described in terms of metric dynamics.

For this purpose, we use the geodesic equation (3.3) and the condition of closed trajectory, eq. (3.4). Closed orbits are stable. For simplicity, we assume a trajectory to be a circle of radius $R$, and the number of revolutions $n_1 = 1$. Then we obtain:

$$\frac{d\vec{v}}{dt} = -\frac{c^2}{2}\nabla \varepsilon_{00} + \nabla(\vec{v}, \frac{c^2}{2}\frac{\partial \varepsilon_{00}}{\partial \vec{v}}) + [\vec{v}, \vec{\Omega}] \tag{7.1}$$

$$2\pi R = nb \tag{7.2}$$

where $b$ is a pitch of the helical trajectory, $\vec{\Omega} \equiv \dfrac{c^2}{2} rot \dfrac{\partial \varepsilon_{00}}{\partial \vec{v}}$, $n$ is a positive integer. Thus, the basis of quantization is the equation (7.2), which has the mathematical nature.

Solving the equation (7.1), we search for $\vec{v}$ in the form

$$\vec{v} = \vec{V} \exp i\{(\vec{k}, \vec{r}) - \omega t\}, \tag{7.3}$$

and for a correction $\varepsilon_{00}$ to the component of the metric tensor in the form

$$\varepsilon_{00} = \frac{1}{2}\frac{v^2}{c^2} = \frac{1}{2}\frac{V^2}{c^2} \exp 2i\{(\vec{k}, \vec{r}) - \omega t\}, \tag{7.4}$$

considering the real parts in both cases. Then, $\nabla \varepsilon_{00} = 2i\vec{k}\varepsilon_{00}$, $\nabla(\vec{v}, \dfrac{c^2}{2}\dfrac{\partial \varepsilon_{00}}{\partial \vec{v}}) = c^2 2i\vec{k}\varepsilon_{00}$, $\vec{\Omega} = \dfrac{i}{2}[\vec{V}, \vec{k}]\exp\{i\{(\vec{k}, \vec{r}) - \omega t\}\}$, and equation (7.1) gives

$$\frac{d\vec{v}}{dt} = ic^2 \{\vec{w} + [\vec{w}, [\vec{w}, \vec{k}]]\}\varepsilon_{00}, \tag{7.5}$$

where $\vec{w} = \dfrac{\vec{V}}{V}$, and the dispersion relation has the form

$$\omega \vec{w} = -\frac{1}{2} V \{\vec{k} + [\vec{w},[\vec{w},\vec{k}]]\} \exp i\{(\vec{k},\vec{r}) - \omega t\} \ . \tag{7.6}$$

The obvious connection between the characteristics of the wave and the helix, gives $b = \frac{2\pi}{k}$, so, we find an estimate for the size of the atom from equation (7.2):

$$R = \frac{n}{k} \ . \tag{7.7}$$

The estimates, that follow from eqs. (7.3, 7.4) in this case, are

$$v = V \cos n \tag{7.8}$$

$$\varepsilon_{00} = \frac{1}{2} \frac{V^2}{c^2} \cos^2 n \ . \tag{7.9}$$

Let us, as usual, assume that the atomic transition from one state to another state, characterized by different numbers $n$, changes the atom's energy. Now the energy is calculated using formulas (5.3, 5.4). Substitute the results obtained in this section into them

$$T = \frac{mv^2}{2}\{\sin^2(\vec{v},\vec{\Omega}) + \cos^2(\vec{v},\vec{\Omega})\} = \frac{mV^2}{2}\cos^2 n \tag{7.10}$$

$$U = \frac{mc^2}{2}\left\{\varepsilon_{00} - (\vec{v}, \frac{\partial \varepsilon_{00}}{\partial \vec{v}}) - \varphi_{(a)}\right\} = \frac{mc^2}{2}(\varepsilon_{00} - 2\varepsilon_{00} + \varepsilon_{00}) = 0 \tag{7.11}$$

It turns out that in this case the motions along the various closed paths are determined by the same (zero) value of the "potential energy". This circumstance arising in metric dynamics is a mathematical reflection of the "physical" Bohr postulate of the existence of stationary orbits, i.e., of the stability of such a dynamic system as an atom, and even in several possible states. Thus,

$$|E_n - E_m| = \frac{mV^2}{2} |\cos^2 n - \cos^2 m| \ . \tag{7.12}$$

If the further interpretation suggests that the transition from one state to another results in the energy difference emission from the system in the form of a quantum, then (7.12) shows that the spectrum of radiation should be almost continuous. Indeed, the difference of the squares of cosines of integers can be made almost any (less than unity) with an appropriate choice of integers. This is consistent with the observable fact that xenon and other light sources based on inert gases provide a substantially continuous spectrum of radiation.

Since Planck's quantum hypothesis is natural for metric dynamics, and the photon energy is $E = h\nu$, let us use the value of $h$, obtained from experiments with radiation, to estimate the rate of finite motion of the particle with regard to $E = |E_n - E_m|$. The frequency of visible light, $\nu \sim 10^{15} s^{-1}$, and electron mass is equal to $m_e \sim 10^{-30} kg$, both measured independently, we obtain $V \sim 10^6 m/s$. Then with the help of (7.6, 7.7), we get an estimate $R \sim \frac{1}{k} \sim \frac{V}{\nu} \sim 10^{-10} m$ for the size of an atom, which coincides with the known one.

The difference between the results of the theory of hydrogen atom proposed by Bohr and based on the analysis of the spectra, and the results obtained here, stems from the fact that a hydrogen atom is described in the framework of the two-body problem with Coulomb interaction potential. However, for more complex situation, the problem of many bodies (with the Coulomb potential) has no analytical solution. So, the use of the classical approach to describe transitions in atoms of inert gases[3], occurring within the "electron shell" containing several bodies, is impossible. This was one of the reasons why the further development of quantum mechanics has followed an abstract and even contradictory way rejecting visual representations.

As it is known, the Schrodinger equation was not derived but postulated by analogy with classical optics, on the basis of generalization of experimental data that might be interpreted sometimes as a manifestation of the corpuscular properties, and sometimes as a manifestation of the wave properties of micro-particles. This made it possible to attach to some extent a physical meaning to the formal analogy between Jacobi theory and the wave theory. The first allows us to calculate the classical action for a particle according to the formula

$$S = Et - mv(\alpha x + \beta y + \sqrt{1 - \alpha^2 - \beta^2}\, z), \qquad (7.13)$$

while the second allows us to calculate the phase of the wave by the formula

$$\varphi = vt - \frac{1}{\lambda}(\alpha x + \beta y + \sqrt{1 - \alpha^2 - \beta^2}\, z). \qquad (7.14)$$

Relating the motion of a particle having energy, $E$ with the propagation of the wave having frequency, $v$, using Planck's constant, $h$, and relating the classical particle momentum with the corresponding wavelength by the formula introduced by de Broglie,

$$\lambda = \frac{h}{mv}, \qquad (7.15)$$

Schrodinger used the expression

$$\psi = a \exp(i\frac{2\pi}{h} S), \qquad (7.16)$$

known as the wave function to describe the dynamics of the micro-system.

There is an important distinction between equations (7.13) and (7.14). It is in the fact that the expression (7.13) for the action is based on the classical equations of mechanics, where the free motion of particles is uniform and rectilinear, and this is a *philosophical assumption*, impossible to verify. And the equation (7.14) follows from the formal wave equation, which proved to be suitable for describing the available *observations*. It is this fact that has led to well-known paradoxes associated with the interpretation of these concepts.

In metrical dynamics with the geodesics (3.1, 3.2) as the equations of motion, the above assumptions are not used, and the trajectory of the free motion is a helix of general form. Therefore, the motion of a particle is naturally characterized by a concept of phase, and plane waves correspond to the projections of the trajectory of motion on various planes[4]. At the same time, the wave properties of the moving micro-particle such as rounding the obstacles and the emergence of a variety of diffraction and interference patterns, depending on the boundary conditions, can be now described by selecting the suitable geometry of modeling space.

Note that for the applications related to the homogeneous Maxwell equations, we can consider a value, $\Psi = E + iH$ similar to the wave function, but expressed in terms of geometric characteristics given in eqs. (6.7), (6.9) and (6.10). Then this pair of Maxwell's equations can be transformed into equation

$$i\frac{\partial \Psi}{\partial t} = c \cdot rot \Psi, \qquad (7.17)$$

which can be used in the problems normally associated with the solution of the Schrodinger equation.

---

[3] And in other complex atoms

[4] The formalism of the regular quantum mechanics corresponds to the description of these very projections.

## 8. Gravitation

In some cases, the astronomical objects and processes observed at the galactic scale and higher allow one to perform remote kinematic measurements. However, everything regarding the related physical, and in particular, gravitational fields, is only a matter of interpretation based on the theory used. Since at the scale of planetary systems the geometric approach of GRT was not only adequate, but has successfully predicted new effects, it became the basis of interpretation on much larger scales. However, this cannot mean that the approach itself as well as Riemann geometry is the only suitable in all cases. Still, the developed alternative theories have not led to significant progress.

Therefore, the approach based on the geometry discussed here was proposed and used as the theory of gravity, and it was natural to call it anisotropic geometrodynamics (AGD). It is described in detail in papers [10, 16] and in the monograph [9]. Here we shall mention only the main points and present the results previously obtained in these studies.

The equation of motion is again determined by the equation of the geodesic eq. (3.3), and the gravitational force in accordance with Proposition 2 has the form

$$\vec{F}_g = \frac{mc^2}{2}\left\{-\nabla \varepsilon_{00} + \nabla(\vec{v}, \frac{\partial \varepsilon_{00}}{\partial \vec{v}}) + [\vec{v}, rot\frac{\partial \varepsilon_{00}}{\partial \vec{v}}]\right\}. \qquad (8.1)$$

It contains the contributions of three terms. If there are no additional assumptions, neither one of the terms can be neglected in comparison with the others. And then the difference with the case of the use of Riemann geometry in GRT is obvious. The physical meaning, which can be attributed to the additional terms in the interpretation, is also clear. Namely, the equivalence principle must take a generalized character: since the force of inertia may depend on the velocity of the body, the force of gravity should also depend on it. For this reason, AGD may also be called a generalized theory of equivalence (GTE). Depending on the angle between the velocity vector of a particle, and vectors associated with $\frac{\partial \varepsilon_{00}}{\partial \vec{v}}$, their contributions, comparable in magnitude, can have different signs. This reflects the possibility to *observe* not only attractive action but also repulsive or tangential action. Using the concepts of "metric field" given in Section 4, one can calculate any model situation, in which the distributions of moving sources $\rho$ and $\vec{j}$ allow to calculate $\vec{F}^{(*)}$ and $\vec{F}^{(**)}$. For a simple system where the test body moves along a closed *current* surrounding a singular *source*, the equation of motion will lead to

$$v^2 = \frac{C_1}{r} \pm vC_2, \qquad (8.2)$$

where $C_1$ and $C_2$ are constants, and the sign depends on the direction of motion of a probe body. Solutions are of the form

$$v = \frac{C_2}{2}(1 - \sqrt{1 \pm \frac{4C_1}{rC_2{}^2}}) \qquad (8.3)$$

$$v = \frac{C_2}{2}(1 + \sqrt{1 \pm \frac{4C_1}{rC_2{}^2}}) \qquad (8.4)$$

If we apply this model to describe the motion of stars in a spiral galaxy, the formula eq. (8.3) corresponds to the Newtonian result and means that the velocity of the orbital motion decreases with increasing distance from the center. This is consistent with general relativity and Newton gravitation. At the same time, formula eq. (8.4) describes the situation when the speed of the orbital motion of a star tends to a constant, which corresponds to the observed flat rotation curves. In [9], [10] and [16] it is shown that there is also a quantitative agreement with observations. Writing down the explicit expressions for the constants $C_1$ and $C_2$ through the parameters of the problem and assuming that the luminosity of a spiral galaxy is proportional to its area, we get an empirical Tully-Fisher law, $v_{orb} \sim L_{lum}{}^{1/4}$, which has no explanation in general relativity.

Thus, the use of anisotropic geometry to describe the astronomical phenomena on the galactic scale makes it possible to adequately describe the observations, and no "dark matter" is required. Other results obtained in the framework of the AGD in [9], [10] and [16] are summarized in the following table:

*Table 1*

| AGD Results | Observations | Modern interpretation |
| --- | --- | --- |
| 1. If the gravitational field does not depend on the velocities of the bodies, the AGD equations become the GRT equations. | Confirm the theory on the scale of the solar system | GRT |
| 2. When a body performs the gravitational acceleration maneuver driving in the planetary system, an additional acceleration directed toward the center, which is proportional to $cH$ should be observed. ($H$ is Hubble constant) | Effect of "Pioneers" which has the order of $cH$. | 14 various explanations that take into account the following factors: technical, data processing, space objects, and «new physics". They are comparable in order of magnitude, which does not allow choosing only one of them. |
| 3. The rotation curves for spiral galaxies should be flat | Yes. | 1. There is dark matter, whose mass is 4 to 7 times the mass of the luminous (baryonic) matter in the galaxy. Dark matter particles possess exotic properties and have not yet been found in a direct experiment<br>2. In the MOND theory the change of the Newtonian dynamics equations is proposed by introducing an additional term, ensuring fit to the observations. |
| 4. The orbital velocities of stars and gas, corresponding to the flat rotation curves of spiral galaxies must comply with centripetal acceleration of order $cH$ at distances of the order of the | Yes. | Interpretation is missing. |
| radius of the galaxy. | | |
| 5. If the luminosity of a galaxy is proportional to its mass, and the mass of a spiral galaxy is distributed in the plane, the luminosity should be proportional to the fourth power of the orbital velocity of the stars on the periphery. | Tully-Fisher law resulting from the observations. | Interpretation is missing. Note that in these observations, the hypothetical dark matter present in a galaxy does not manifest itself, which contradicts its supposed property to have a gravitational action. |
| 6. The galaxies with large angular momentum should have arms. | Yes, spiral galaxies. | The theory of density waves. It does not predict the bars. |
| 7. Spiral galaxies should have bars. | Yes. | Interpretation is missing. |
| 8. The motion of individual objects in the plane of the spiral galaxy must contradict the Kepler law | Observations of globular clusters in the galactic plane disclose a violation of the Kepler law statistics: the number of clusters in the center of the galaxy is significantly greater than in the periphery. | Interpretation is missing. |
| 9. The motion of objects in the plane of a spiral galaxy and in the plane perpendicular to it should be different | Observations of globular clusters in the perpendicular plane correspond to Kepler's law as opposed to item 8 | Interpretation is missing. |
| 10. In some gravitational lenses having the necessary orientation, there must be a significant excess of refraction compared with the estimate following from GRT | Yes. | Considered to be related to the action of dark matter and is used to estimate its amount. |
| 11. Gravitational lenses, which are spiral galaxies, with the profile orientation should give the asymmetry in the image. | Yes. For example, the Einstein Cross. | Interpretation is missing. |
| 12. In view of mass and energy equivalence, the clusters of galaxies should have larger mass than it can be assessed by their luminosity, and larger than a correction, which follows from the GRT. | F. Zwicky observations. | "Hidden mass" (in accord with F.Zwicky) and dark matter. |
| 13. In collisions of individual galaxies, there should be manifested the excess of "mass-energy" associated with mutual movement | Observation of the collision of galaxies in the Bullet cluster made by "Chandra" observatory. | Dark matter. |
| 14. The red shift in the emission of distant objects should increase linearly with distance, which is associated with vortex motion of object of the scale of galaxies and higher. | Empirical law discovered by E.Hubble | Cosmological expansion of the universe |
| 15. There can exist concave gravitational lens, resulting in incorrect (overvalued) determination of the distance to the corresponding light sources. | Deviations from the linear Hubble law, relevant to this hypothesis were detected. | Expansion of the universe is accelerated due to the dark energy (of repulsion). |
| 16. The distribution of matter near the nuclei of spiral galaxies can have a characteristic form (the "infinity" sign) | Discovered by "Herschel" observatory under the supervision of cold gas clouds in the center of our galaxy. | Interpretation is missing. |

Additionally, one can mention the recent precision measurements of the motion of geodetic satellites "Lageos" [17]. The authors interpret the results in the classical spirit as the influence of "ether". Using this concept makes one return to the experimental problems of ether detection (similarly to the case of the experimental discovery of dark matter carriers). However, these observations can also be described in terms of the AGD as a particular case of the metric dynamics.

**Discussion**

The transition from the dynamic to the geometric formulation of the laws of mechanics refers to the ideas of E.Mach - one of the most famous physicists and philosophers of those who tried to revise some inconsistent ideas of classical mechanics. The problem of the total interrelationship, raised by him, is reflected in the choice of appropriate geometry to describe a given class of phenomena, while the problem of inertia is withdrawn in light of Proposition 2. At the same time, such concepts as energy, momentum, closed system change their meaning, since we stop the search for hypothetical reasons of motion (or physical properties of the world around), and consider only the necessary geometrical properties of modeling space (mathematical description), that are inseparably connected with the motion of a probe body and the distribution of moving sources.

Thus, under metric dynamics, it is again proposed to separate the corpuscular and wave properties of the observed phenomena, as Madelung and de Broglie tried to do. However, now there is no need to introduce the material carrier of these properties, such as ether, as it was assumed in Maxwell-Lorentz theory, and one can just choose the proper geometry of modeling space, as it was proposed by Minkowski.

The point of this approach is an effort to separate explicitly the physical and mathematical worlds in the spirit of R. Penrose ideas [19] on their objective existence. In other words, it is proposed to avoid using the tools (mathematics) as an object of the outside world ready for the study and measurement (physics). In fact, the question is formulated as follows: provide such a way to calculate the distance with the help of measured coordinates that makes the observed trajectory coincide with the geodesic of the corresponding space. This means the choice of geometry. The necessity of this approach stems from the fact that at mega- and micro-scales, the possibility to directly measure distances disappears. All of them are determined by measuring the time of propagation of standardized signals (e.g., photons) in an environment where the objects of measurement are present. It is impossible to measure directly how the processes under study flow, and all one can do is to produce a self-consistent description.

The question of which property of the object nature does the wave function correspond to, and what does its reduction mean, has generated the debate among physicists and philosophers that lasted for several decades. Initially, E.Madelung proposed the hydrodynamic interpretation of quantum mechanics, which hinted at a return to the ether. Then in 1927, de Broglie tried to divide the wave and particle, introducing an idea of the pilot wave. However, under the pressure of the universally recognized authority, preferring to speak of "probability waves", that, incidentally, excluded the concept of trajectory from consideration, de Broglie gave up the idea. Nevertheless, in 1951 he wrote the following observation published only after 30 years [16]:

"A*pparently, there is no mechanism based on classical or relativistic notions of space and time that can explain such an instantaneous contraction* [reduction of the wave function – S.S.]*, which, however, is closely related to the indivisible nature of the particle. Taking relativistic notions of space and time, Einstein considered this output as an objection to the wave mechanics. Now, when new ideas appear more established, there is an obvious necessity to consider this conclusion as an indication of the inadequacy of our notions of space and time, even if refined in the theory of relativity.* "

And also in [16]:

"*Corpuscular notions make it impossible to introduce the concept of the phase difference.*"

The last remark refers to the key point of the quantum theory regarding the wave properties of micro-particles. The observation of electron beam diffraction on a crystal performed by K.Davisson and L.Germer in 1927, and observations of single electron diffraction performed by I.A.Fabrikant in 1948 (and their subsequent refinements were interpreted as manifestations of the wave properties of micro-particles. And the physical community reconciled with the presence of two mutually exclusive properties in one object, since the calculations based of this assumption were in accord with experiment[5].

In metric dynamics, it is again proposed to divide corpuscular and wave properties of the observed phenomena as Madelung and de Broglie tried to do. However, now there is no need to introduce the material carrier of these properties, such as ether, as assumed in Maxwell-Lorentz theory, but one just choose the proper geometry of modeling space as it was suggested by Minkowski.

From a formal point of view, the use of force fields is equivalent to the use of appropriate space geometry selected for the simulation of physical reality. On the one hand, it returns to the old philosophical debate about the materiality of the field and about the interaction at a distance. As before, it may be subject to personal preferences of the researcher. But just as before, the involvement of new mathematical apparatus may enable a new theoretical progress and compare the results with the results of appropriate experiments.

---

[5] This humility is illustrated by the expressive principle "shut up and calculate".